# Fundamentals of polaritons in strongly anisotropic thin crystal layers

Kirill V. Voronin[1], Gonzalo Álvarez-Pérez[2,3], Christian Lanza[2], Pablo Alonso-González[2,3,†], Alexey Y. Nikitin[1,4,‡]

[1]Donostia International Physics Center (DIPC), Donostia-San Sebastián 20018, Spain.
[2]Department of Physics, University of Oviedo, Oviedo 33006, Spain.
[3]Center of Research on Nanomaterials and Nanotechnology, CINN (CSIC-Universidad de Oviedo), El Entrego 33940, Spain.
[4]IKERBASQUE, Basque Foundation for Science, Bilbao 48013, Spain.

† Corresponding author: pabloalonso@uniovi.es
‡ Corresponding author: alexey@dipc.org

**Abstract.** Polaritons in strongly anisotropic thin layers have recently captured the attention in nanophotonics because of their directional propagation at the nanoscale, which offers unique possibilities for nanooptical applications. However, exploiting the full potential of anisotropic polaritons requires a thorough understanding of their properties, including field confinement, energy and phase propagation direction and losses. Here we fill this critical gap by providing fundamental insights into the propagation of anisotropic polaritons in thin biaxial layers. In particular, we introduce a novel methodology that allows us to represent isofrequency curves of polaritons in strongly anisotropic materials considering that the real and imaginary parts of the wavevector are not parallel. In fact, we analytically show that the direction of the imaginary part of the wavevector is parallel to the group velocity, which can have different, even perpendicular or opposite, directions with respect to the phase velocity. This finding is crucial for understanding polaritonic phenomena in anisotropic media, yet it has so far been widely overlooked in the literature. Additionally, we introduce a criterion for classifying the polaritonic modes in biaxial layers into volume and surface categories, and analyze their dispersion, field structure, and losses. Finally, we discover the existence of anisotropic transverse electric modes, which can exhibit natural canalization. Taken together, our results shed light on hitherto unexplored areas of the theory of electromagnetic modes in thin biaxial layers. Although exemplified for van der Waals α-$MoO_3$ layers, our findings are general for polaritons in other strongly anisotropic biaxial hyperbolic crystals.

## Introduction

Recent experiments have reported the existence of hyperbolic polaritons in strongly anisotropic materials in which the main components of the permittivity tensor have different signs[1-5], which include the van der Waals (vdW) crystals h-BN[1,6], α-$MoO_3$[3,4,7], α-$V_2O_5$[8], $WTe_2$[9], ZrSiSe[10], and the polar dielectrics calcite[11], β-$Ga_2O_3$ (bGO)[12-14] and $CdWO_4$[15]. These exotic light-matter excitations exhibit a number of striking optical effects, such as extreme light confinement[16,17], flat dispersion[18-20], negative phase velocity,[21] anomalous diffraction and focusing[2,16,22-25], negative refraction and reflection[22,26-28], and topological transitions of the dispersion surfaces[18,29,30], making them ideal for various nanophotonic applications. Given the pivotal importance of anisotropy in these excitations, a thorough understanding and precise characterization of their properties are of the highest relevance, specifically concerning the interpretation of near-field optical experiments. Particularly, despite the development of several analytical models that allow obtaining the dispersion relation of anisotropic polaritonic modes, such as approximating a thin layer by a two-dimensional conducting sheet[3,31] or considering much larger momenta than that of free-space light[32], a general analytical tool for the description of polaritons in strongly anisotropic media, such as biaxial layers, has been lacking until now. Although it has been known that in anisotropic media the real and imaginary parts of the momenta of electromagnetic waves are not collinear, in all reported studies on anisotropic polaritons, the electromagnetic fields have been assumed to decay along the wavevector direction. Since the wavevector and propagation directions are generally different, this assumption substantially enlarges the error in estimating the propagation length and the wavelength of the anisotropic polaritonic waves. Furthermore, although the dispersion and field distribution of eigen modes have been studied in uniaxial layers, ribbons and antennas, like hBN[33-35], there is still a lack of understanding of the role that the individual modes play in the propagation of anisotropic polaritons, particularly in systems with low optical symmetry, such as biaxial layers. On the other hand, because of the limitations of current theoretical approximations, transverse electric (TE) polaritonic modes, which can exist in strongly anisotropic media, have not been studied so far.

Here we perform an in-depth theoretical analysis of polaritonic waveguiding modes in thin biaxial layers surrounded by isotropic dielectrics, using α-$MoO_3$ as a representative example. We revisit the calculation of the dispersion relation and the isofrequency curves (IFCs) for the case of non-negligible losses, highlighting that the real and imaginary parts of the polariton wavevector are not parallel in general. In addition, we derive a rigorous procedure to represent the direction of the real part of the polariton wavevector as well as the direction of the imaginary part. We prove that for low losses the direction of the imaginary part of the wavevector coincides to the group velocity, and thus to the propagation direction of the polaritonic wave. Furthermore, we simulate the field distributions of the discrete polariton modes of the thin layer, and provide insights into their relative role on the fields excited by a point dipole source placed above it. We analyze the field confinement of these modes, providing a precise criterion to analytically elucidate their surface- or volume-like character. Finally, we unveil the existence of previously overlooked anisotropic transverse electric (TE) modes propagating along a thin biaxial layer, which exhibit low losses and weak confinement as extracted from our theory. Furthermore, we demonstrate that they can show canalization, i.e., unidirectional propagation, within hyperbolic spectral ranges.

## Results and discussion

We consider an α-$MoO_3$ thin layer of thickness $d$ and dielectric permittivity tensor $\hat{\varepsilon}$ placed in between two isotropic dielectric half-spaces, as shown in Fig. 1A. One of the main axes of the biaxial crystal, [010], is perpendicular to the layer surfaces. Without loss of generality, we choose the coordinate system in which the dielectric permittivity tensor is diagonal $\hat{\varepsilon} = \mathrm{diag}\left[\varepsilon_x, \varepsilon_y, \varepsilon_z\right]$, so that the coordinates $x, y$ and $z$ are collinear to the main crystal axes $[100]$, $[001]$, and $[010]$, respectively. Figure 1B represents the real parts of $\varepsilon_x, \varepsilon_y$ and $\varepsilon_z$ as a function of frequency in the long-wave-infrared (LWIR)[7] and upper far-infrared frequency ranges.[36] α-$MoO_3$ has several reststrahlen bands (RBs) characterized by a negative sign of at least one of the diagonal elements of the tensor $\mathrm{Re}(\hat{\varepsilon})$. In these frequency ranges, the biaxial thin layer supports phonon polaritons (PhPs) with anisotropic dispersion, so that, the wavevector of these modes depends on the propagation direction. Furthermore, within each RB, and depending upon the signs of the elements of $\mathrm{Re}(\hat{\varepsilon})$, PhPs can have either elliptic- or hyperbolic-type dispersion.[3,29] The elliptic-type dispersion is characterized by propagation along all in-plane directions and takes place when the in-plane components of $\mathrm{Re}(\hat{\varepsilon})$ have the same signs. In contrast, PhPs with hyperbolic dispersion can propagate only within an angular sector around one of the main axes. Hyperbolic dispersion takes place in spectral ranges in which $\mathrm{Re}(\varepsilon_x)$ and $\mathrm{Re}(\varepsilon_y)$ have different signs. The propagation of anisotropic polaritons has been consistently analyzed in terms of either the analytical dispersion of TM modes propagating along the two in-plane crystal axes or transfer matrix methods (usually visualized via colorplots showing the poles of the transmission/reflection coefficients)[37]. We have previously derived the explicit expression for the dispersion relation under the assumption that the polaritonic momentum is much larger than the momentum of free-space light, i.e., for highly confined waves, from now on referred to as the high-$k$ approximation[32]:

$$k = \frac{\rho}{d}\left[\operatorname{atan}\left(\frac{\rho\varepsilon_1}{\varepsilon_z}\right) + \operatorname{atan}\left(\frac{\rho\varepsilon_2}{\varepsilon_z}\right) + \pi l\right], \qquad l = 0,1,\dots \tag{1}$$

Here, $k$ is the radial component of the wavevector $\mathbf{k}$; $l$ is the mode number; $\varepsilon_1$ and $\varepsilon_2$ are dielectric permittivities of the superstrate and the substrate, respectively; $\rho = i\sqrt{\frac{\varepsilon_z}{\varepsilon_x\cos^2\alpha+\varepsilon_y\sin^2\alpha}}$, and $\alpha$ is the polar angle between $\mathbf{k}$ and the $x$-axis. Eq. (1) has been repeatedly proven to successfully account for the propagation of highly-confined polaritons in anisotropic thin layers, as illustrated in Figure 1C by a perfect agreement with transfer matrix calculations (dashed lines vs colorplots, respectively).

Note, that the dispersion relation given by Eq. (1) describes polaritons of only one of the two possible polarizations. Indeed, isotropic media supports two orthogonally polarized waves, transverse electric (TE), or s-polarization, and transverse magnetic (TM), or p-polarization, whereas, inside uniaxial media, they are generally referred to as ordinary (exhibiting the same dispersion as isotropic) and extraordinary. However, in biaxial media both waves are extraordinary, but they reduce to ordinary and extraordinary in the uniaxial limit and thus we refer to them as "o" and "e". As demonstrated in Section 1 of the Supporting Information, the electromagnetic fields of the modes inside a biaxial layer have predominantly “e”-polarization. In turn, inside the isotropic dielectric half-spaces, the mode has p-polarization (TM mode). Importantly, neither the fields of the “o”-polarization inside the crystal layer nor the s-polarized fields (TE modes) inside the isotropic dielectric half-spaces are captured by Eq. (1). Therefore, the description of these polaritonic modes requires other analytical approaches, which we discuss below in detail. For simplicity, from now on we will refer to the polarization of the electromagnetic modes with respect to their electric and magnetic fields outside the biaxial thin layer, that is, TM and TE modes.

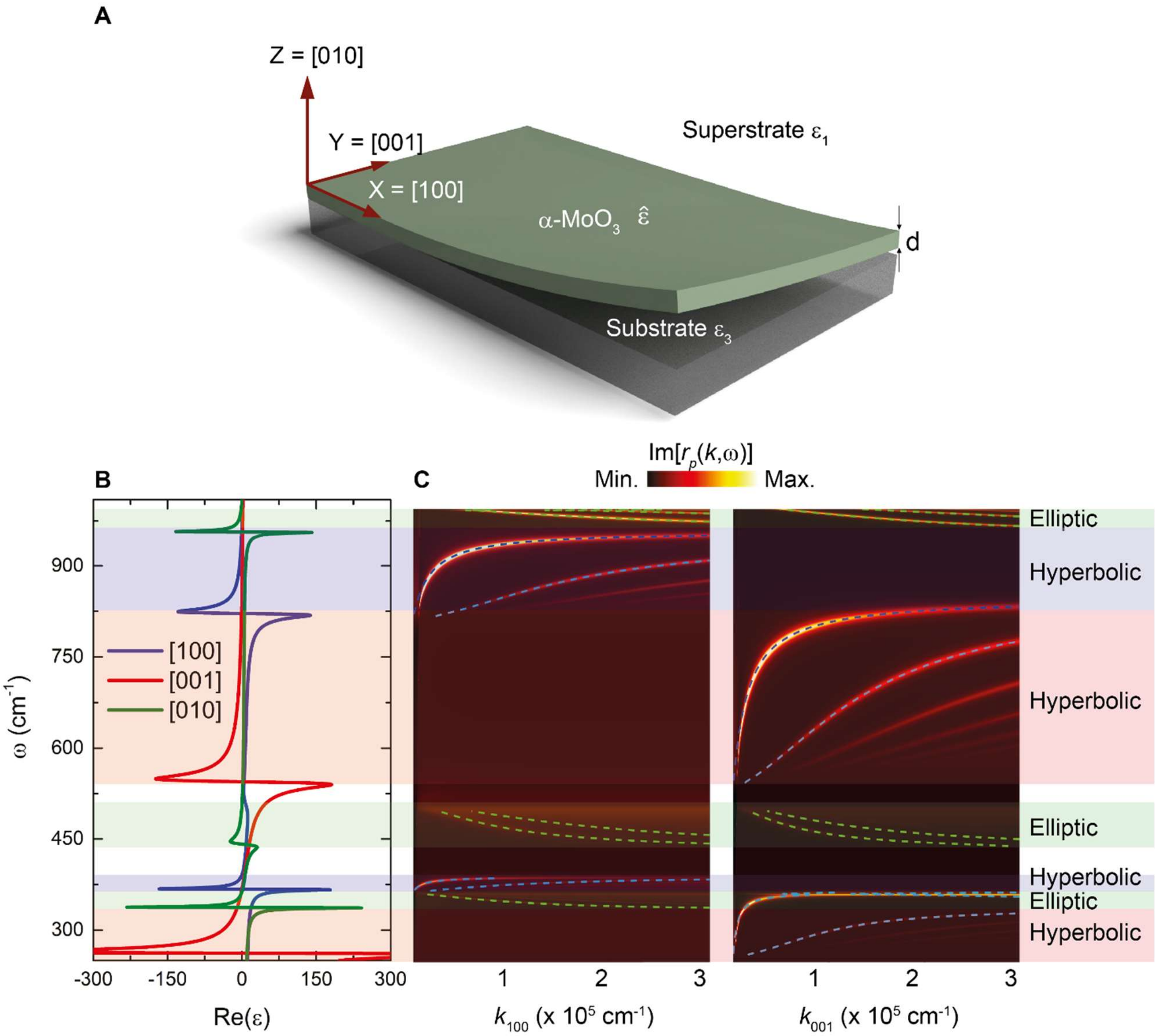

**Figure 1. Dispersion of polaritons in a biaxial crystal layer (α-$MoO_3$) A)** Schematic of the material structure: a layer of α-$MoO_3$ with thickness $d = 500\ \mathrm{nm}$ and dielectric permittivity tensor $\hat{\varepsilon}$ is placed on an isotropic substrate with dielectric permittivity $\varepsilon_3$. The layer is covered by an isotropic medium with dielectric permittivity $\varepsilon_1$. The coordinate system $(x, y, z)$ is aligned along the main axes of the permittivity tensor, which are parallel to the crystallographic axes [100], [001], and [010]. **B)** Real part of the main components of the dielectric permittivity tensor as a function of frequency. Colored background regions indicate different reststrahlen bands (RBs) in which in-plane elliptic and in-plane hyperbolic polaritons are supported. **C)** Transfer matrix (colorplot) and analytic calculations (dashed curves) of the dispersion relation of TM-polarized PhPs modes propagating along the main axes of the biaxial layer. Frequencies within different RBs are considered. A perfect concordance between the transfer matrix calculations and the analytic results is obtained, emphasizing the wide applicability of our simplified expression for the dispersion relation.

Apart from only describing the TM modes, Eq. (1) also hides some important effects appearing when losses inside the layer become non-negligible. This is particularly dramatic for hyperbolic polaritons with their large wavevectors $\mathbf{k}$ (both their real and imaginary parts) pointing towards the asymptotes of the IFC. Namely, Eq. (1) in its explicit shape assumes that the real and imaginary parts of the complex-valued wavevector $\mathbf{k}$ are collinear, so that their directions are both defined by a single angle $\alpha$ with respect to the $x$-axis. Such collinearity indeed takes place in isotropic media, or in the case of polariton propagation along the main axes of an in-plane anisotropic thin layer ($\alpha = 0$ or $\alpha = \pi/2$, depending on the RB). However, for an arbitrary in-plane direction along an anisotropic thin layer, the orientations of the real and imaginary parts of $\mathbf{k}$ can significantly differ. Generally, if $\mathbf{k} = \mathbf{k}' + i\mathbf{k}''$, where $\mathbf{k}' = \mathrm{Re}(\mathbf{k})$ and $\mathbf{k}'' = \mathrm{Im}(\mathbf{k})$, then $\mathbf{k}'$ is not parallel to $\mathbf{k}''$. To gain insights into the non-collinearity between the real and imaginary parts of $\mathbf{k}$, we have to rewrite Eq. (1) in an implicit form as $f(k_x, k_y, \omega) = 0$, where $f(k_x, k_y, \omega)$ is given by

$$f(k_x, k_y, \omega) = \sqrt{k_x^2 + k_y^2} - \frac{\rho(k_x,k_y,\omega)}{d}\left[\mathrm{atan}\left(\frac{\rho(k_x,k_y,\omega)\varepsilon_1}{\varepsilon_z}\right) + \mathrm{atan}\left(\frac{\rho(k_x,k_y,\omega)\varepsilon_2}{\varepsilon_z}\right) + \pi l\right], \quad (2)$$

and where $\rho(k_x, k_y, \omega) = i\sqrt{\frac{\varepsilon_z(k_x^2+k_y^2)}{\varepsilon_x k_x^2+\varepsilon_y k_y^2}}$ with all the dielectric permittivities being functions of $\omega$ and $k_x$ and $k_y$ being complex-valued projections of $\mathbf{k}$ onto $x$- and $y$-axes, $\mathbf{k} = (k_x, \quad k_y)$.

When the dielectric permittivities are real-valued, Eq. (2) also has real-valued solutions for $k_x$, $k_y$, and $\omega$. These solutions can be represented by a dispersion surface in the three-dimensional space $(k_x, k_y, \omega)$. The cross-section of the dispersion surface by the plane of a constant frequency represents a two-dimensional isofrequency curve (IFC). However, in realistic materials, dielectric permittivities are complex-valued, so both momentum and frequency are complex-valued as well, i.e., $k_x = k_x' + ik_x''$, $k_y = k_y' + ik_y''$, and $\omega = \omega' + i\omega''$. Then, Eq. (2) for three complex-valued unknowns, splits into a system of two equations, $\mathrm{Re}[f(k_x, k_y, \omega)] = 0$ and $\mathrm{Im}[f(k_x, k_y, \omega)] = 0$, for six real-valued unknowns, $k_x'$, $k_x''$, $k_y'$, $k_y''$, $\omega'$, $\omega''$. The solutions of these equations form a four-dimensional hypersurface in a six-dimensional space. For the visualization and interpretation of the generalized dispersion, we have developed a novel procedure based on the minimization of the losses $\mathbf{k}''$ for each direction of $\mathbf{k}'$ (see Supporting Information, section 2). It consists of considering a continuous monochromatic illumination (real-valued $\omega$) and splitting the dispersion hypersurface into two dispersion surfaces, for the real and imaginary components of $\mathbf{k}$. Following this procedure, we represent in Fig. 2 the IFCs (cross-sections of these surfaces at certain frequencies) of the polaritonic modes excited in an α-$MoO_3$ thin layer at frequencies residing in two different RBs, one hyperbolic ($\omega = 955\ \mathrm{cm}^{-1}$) and one elliptic ($\omega = 980\ \mathrm{cm}^{-1}$). Figure 2A (top panel) shows the IFCs in the hyperbolic range (the real and imaginary parts of $\mathbf{k}$ are shown by black and red curves, respectively). To clearly visualize the confinement degree of the modes, we use the normalized wavevector $\mathbf{q} = \mathbf{k}/k_0$, where $k_0 = \frac{\omega}{c}$, with $\omega$ and $c$ being the frequency and the speed of light, respectively. For an easier visual correlation between the real IFC (IFCR) and the imaginary IFC (IFCI), we select several polaritonic modes (plane waves with given momenta) with different directions of $\mathbf{q}'$ labeled with numbers, (the same number in both IFCs represents $\mathbf{q}'$ and $\mathbf{q}''$ of the same plane wave). For example, the mode labeled by "1" propagates along the [100] axis and has collinear $\mathbf{q}'$ and $\mathbf{q}''$, similar to a plane wave in an isotropic medium. In contrast, the mode labeled by "4", for which $\mathbf{q}'$ takes values close to the asymptote of the hyperbola-like IFCR, exhibits almost perpendicular $\mathbf{q}'$ and $\mathbf{q}''$. A rather different situation is obtained within the elliptic RB at $\omega = 980\ \mathrm{cm}^{-1}$. Figure 2B (top panel) shows the IFCR and IFCI showing polaritonic modes with almost opposite directions of $\mathbf{q}'$ and $\mathbf{q}''$, even when $\mathbf{q}'$ is oriented along the main in-plane crystal axes. Such anti-orientation of $\mathbf{q}'$ and $\mathbf{q}''$ indicates a negative sign of the phase velocity.[3,4,21] The minimization procedure thus allows us to calculate the real and imaginary parts of $\mathbf{q}$ for polaritonic modes contributing to the field distribution far from the excitation source without assuming collinearity between them.

For a deeper understanding of the propagation properties of in-plane anisotropic polaritons, it is fundamental to consider the relationship between the directions of the group velocity and the imaginary part of $\mathbf{q}$. To do this, we consider the case of relatively small losses, in which the group velocity is well-defined and its modulus and direction correspond to the propagation velocity and direction of the wave packet, respectively. Using the standard expression for the group velocity and assuming that $\varepsilon_i'' \ll \varepsilon_i'$ and $q'' \ll q'$, we obtain that the condition for minimizing the imaginary part of $\mathbf{q}$ requires the collinearity of $\mathbf{v}_{gr}$ and $\mathbf{q}''$ (see Section 3 in the Supporting Information for more details). Note that the direction of $\mathbf{v}_{gr}$ is usually assumed to be equal to the direction of the Poynting vector $\mathbf{S}$. This is proven only for an unbounded anisotropic[38,39] medium, or even more general, for a bi-gyrotropic medium[40]. However, for the in-plane Poynting vector of anisotropic polaritons propagating along biaxial layers, the proof of this statement is an open problem. Indeed, despite the numerical methods that allow to calculate the Poynting vector in each point of a layered structure,[41] the analytical expression for the direction of the Poynting vector has not been derived yet. Moreover, the direction of the Poynting vector is different in different planes parallel to the layer surfaces, which, therefore, implies that to introduce the in-plane Poynting vector it is needed to at least perform an averaging along the z-axis. Consequently, the Poynting vector and energy transfer in biaxial layers is an important mathematical problem for further research beyond the present work.

To illustrate the non-collinearity of the real and imaginary parts of $\mathbf{q}$ for polaritons in anisotropic media, we represent in the bottom panels of Figs. 2A, B the spatial distribution of the vertical component of the electric field, $\mathrm{Re}(E_z)$, for the very different polaritonic modes M0 (fundamental l=0 mode) and M1 (first higher order l=1 mode) in an α-$MoO_3$ thin layer. In Fig. 2A the M0 mode exhibits almost orthogonal $\mathbf{q}'$ and $\mathbf{q}''$ (top panel) indicating that the direction of the amplitude decay and the direction of the phase velocity are almost perpendicular to each other (bottom panel). In Fig. 2B the M1 mode exhibits almost oppositely oriented $\mathbf{q}'$ and $\mathbf{q}''$ (top panel) indicating a negative phase velocity (bottom panel). The IFCIs, obtained by minimizing the losses (see Section2 of the Supporting Information), differ significantly from those obtained by assuming parallel $\mathbf{q}'$ and $\mathbf{q}''$ (see Section 4 in the Supporting Information), showcasing the importance of our procedure to get accurate results and predictions over the propagation direction and decay length of hyperbolic polaritons. Interestingly, in the elliptic RB (Fig. 2B), the IFCRs obtained by both methods are similar, which is explained by the fact that $q'' \ll q'$, and, therefore, the assumption of parallel $\mathbf{q}'$ and $\mathbf{q}''$ provides a sufficiently accurate result for $\mathbf{q}'$. In contrast, in the hyperbolic RB, the IFCRs are similar only near the vertex of the hyperbolic curve, deviating strongly near the asymptotes. This can be explained by the fact that near the asymptotes the angle between the real part of $\mathbf{q}$ and the group velocity tends to $\pi/2$, and therefore, $\mathbf{q}'' \to \infty$ (see Section 3 in the Supporting Information), which breaks the condition $q'' \ll q'$ and makes $q'$ to depend significantly on $q''$. Consequently, the assumption that $\mathbf{q}'$ and $\mathbf{q}''$ are parallel is evidently unsuitable at these points.

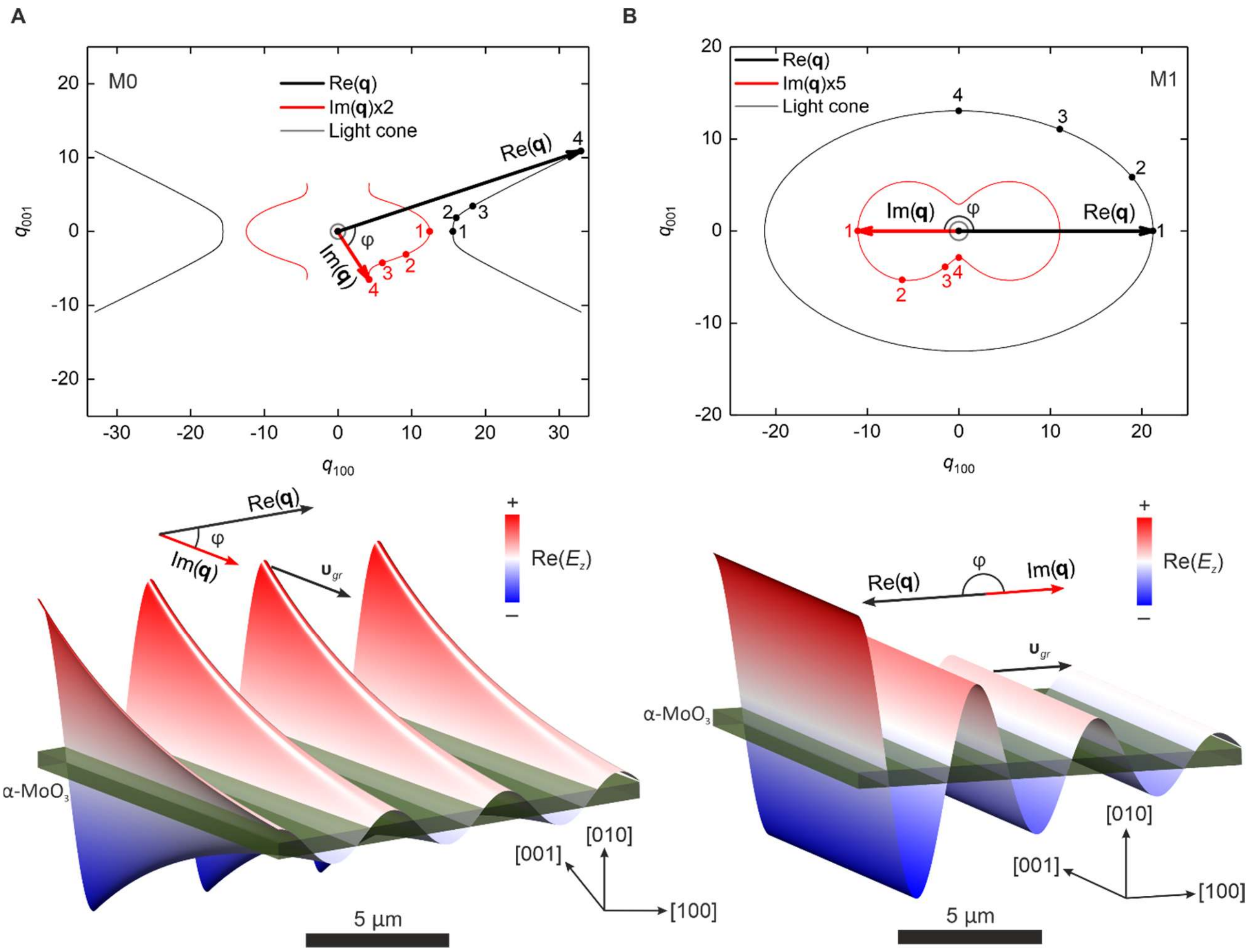

**Figure 2. Fundamental PhPs modes in a free-standing α-MoO₃ layer. A) Top**: IFCs for the $\mathbf{q}'$ (black curve) and $\mathbf{q}''$ (red curve) for the $l = 0$ mode (M0) within the hyperbolic RB ($\omega = 955\ \mathrm{cm}^{-1}$). The IFCs are found under the condition of minimal $\mathbf{q}''$ for every given direction of $\mathbf{q}'$, yielding the mode dominating in the field from a point source. Since the real and imaginary parts of $\mathbf{q}$ are not parallel, we label different points on IFCs for $\mathbf{q}'$ and $\mathbf{q}''$ by numbers, to illustrate that they all characterize the same plane waves. Note, that $\mathbf{q}''$ was increased by a factor of two for a better visualization of the curve. **Bottom**: Schematic of the real part of the vertical component of the electric field, $\mathrm{Re}(E_z)$, for the M0 mode when $\mathbf{q}'$ points towards the asymptote of the IFC and thus with $\mathbf{q}''$ almost perpendicular to it. **B)** Same than (A) for the $l = 1$ mode (M1) within the elliptic RB ($\omega = 980\ \mathrm{cm}^{-1}$). In this case, the real and imaginary parts of $\mathbf{q}$ are oppositely directed, revealing a negative phase velocity. Note, that $\mathbf{q}''$ was increased by a factor of five for a better visualization of the curve. A thickness of the biaxial layer of 500 nm is considered in all the calculations.

To visualize the excitation of the TM polaritonic modes in α-$MoO_3$ thin layers, we perform full-wave electromagnetic simulations using a finite element method (FEM). In particular, we simulate the electric field of the polaritons excited by a vertically oriented electric dipole source, which represents the simplest model of the s-SNOM tip typically employed in near-field polariton interferometry experiments. A snapshot of the electric field at a frequency $\omega = 930\ \mathrm{cm}^{-1}$ is shown by the colorplot in Fig. 3A. At this frequency, the out-of-plane component of the dielectric permittivity tensor is positive ($\mathrm{Re}(\varepsilon_{010}) > 0$), while the in-plane components have different signs, specifically, $\mathrm{Re}(\varepsilon_{100}) < 0$ and $\mathrm{Re}(\varepsilon_{001}) > 0$. One can clearly recognize field oscillations with concave wavefronts, revealing polaritons with hyperbola-like IFCR (see inset in Fig. 3A showing the IFCR of the fundamental M0 mode). The period of the oscillations along the [100] axis is 2.80 µm, which is consistent with the dispersion relation given by Eq. (2) for $l$ =0 that predicts $2\pi/k_{100} = 2.86\ \mu\mathrm{m}$. This excellent agreement indicates that the contribution of the M0 mode along the [100] direction is dominating.

To gain further insights into the polaritons generated by a point (dipolar) source, we analyze in more detail the spatial distribution of their electric field, $\mathrm{Re}(E_z)$. To that end, we plot in Figs. 3B, C (left panel) the field distribution $\mathrm{Re}(E_z)$ generated by a vertical electric dipole (placed at $x = 0$, $y = 0$) in the $uz$ and $vz$ planes, where the directions *u* and *v* are shown in Fig 3A by black arrows. To interpret the obtained field distribution, we suppose that at any point on a plane parallel to the biaxial layer, the in-plane components of the group velocity and imaginary part of the wavevector are aligned with the radius vector. The latter is directed from the origin $x = 0$, $y = 0$ to the observation point. Therefore, one can expect that the field distribution along any direction is dominated by the field of the mode with the imaginary part of its wavevector parallel to the radius vector. In particular, the contribution of the polaritons with the wavevectors $\mathbf{q}_\mathrm{u}$ and $\mathbf{q}_\mathrm{v}$ (see IFCR in the inset of Fig. 3A) dominates along the *u* and *v* directions, respectively. As a validation, we compare the field distributions shown in Figs. 3B, C (left panel) with the analytically calculated field $\mathrm{Re}(E_z)$ (see Section 5 in the Supporting Information) for the M0 mode with wavevectors $\mathbf{q}_\mathrm{u}$ and $\mathbf{q}_\mathrm{v}$ (Figs. 3B, C, middle panel). Along the main axis *u*, both real and imaginary parts of the wavevector are parallel. For this reason, according to Fig. 3B, the polariton wavelength in both the left and middle panels, as well as the vertical confinement of the electric field, show a perfect agreement, thus verifying our assumption on the dominating role of the M0 mode to the field created by the dipole. In a large contrast, along the *v*-axis the real and imaginary parts of the wavevector are not parallel, i.e., $\mathbf{q}'_\mathrm{v} \nparallel \mathbf{q}''_\mathrm{v}$. On the other hand, the mode propagating along the *v* direction with wavevector $\mathbf{q}'_\mathrm{v}$ has a group velocity $\boldsymbol{\upsilon}_{gr}$, which is parallel to $\mathbf{q}''_\mathrm{v}$ and aligned along the *v*-axis (inset to Fig 3A). Consequently, (see details in Section 5 of the Supporting information), the electric field along the *v* direction depends upon the distance $r$ as $E_z \sim e^{ik_0 q'_v r \cos\varphi_\mathrm{v} - k_0 q''_v r}$, where $\varphi_\mathrm{v} \approx 1\ \mathrm{rad}$ is the angle between $\mathbf{q}''_\mathrm{v}$ and $\mathbf{q}'_\mathrm{v}$. In the middle panel of Fig. 3C, we plot the analytically calculated field $\mathrm{Re}(E_z)$ for the mode with wavevector $\mathbf{q}_\mathrm{v}$. The distances between the neighboring maxima/minima of $\mathrm{Re}(E_z)$ in the left and middle panels, as well as the vertical confinement of the electric field, show a perfect agreement, clearly indicating the dominating role of the M0 mode with $\mathbf{q}''_\mathrm{v}$ parallel to the *v* direction in the field distribution created by the dipole. This perfect agreement between the field distribution created by the dipole place over the surface of a biaxial layer along any in-plane direction and the field of the mode with the group velocity and imaginary part of the wavevector parallel to this direction showcases once more the accuracy and relevance of our theoretical approach to predict the polaritonic properties of thin anisotropic (biaxial) layers. Importantly, our analysis becomes particularly relevant when polaritons propagate along directions that are not aligned with the main crystal axes. Polaritonic propagation along these directions plays a crucial role in the demonstration and exploitation of exotic optical phenomena, such as negative refraction[22,27,28], reflection[26], or hyper-focusing.[2,16,23]

Now, we examine how in-plane anisotropic polaritons propagate in a thin layer when the out-of-plane component of the dielectric permittivity tensor is negative, i.e., $\mathrm{Re}(\varepsilon_{010}) < 0$. To do so, we plot in Fig. 3D a snapshot of the electric field at $\omega = 355\ \mathrm{cm}^{-1}$ in the $xy$ plane 50 nm above the 500 nm-thick α-$MoO_3$ layer. At this lower frequency, the in-plane components of the dielectric permittivity tensor have different signs: $\mathrm{Re}(\varepsilon_{100}) > 0$, $\mathrm{Re}(\varepsilon_{001}) < 0$. The resulting field distribution shows that the polaritons propagate along all in-plane directions, in contrast to the typical propagation of anisotropic polaritons in spectral regions where the out-of-plane component of the permittivity is positive (see e.g., Fig. 3A). The period of the oscillations along the *u*-axis (Fig. 3E, left panel) is 5.20 μm, thus being consistent with the dispersion relation (2) for $l = 0$, i.e., the M0 mode, which yields a period $2\pi/k_{100} = 5.06\ \mu\mathrm{m}$. But, strikingly, the period of the oscillations along the $w$-axis is 3 µm (Fig. 3F, left panel), consistent with the dispersion relation (2) for $l = 1$, i.e., the M1 mode, which yields a period $2\pi/k_{001} = 2.86\ \mu\mathrm{m}$. To further corroborate the contribution of the M0 and M1 modes into the electromagnetic field created by a point source, we compare the $\mathrm{Re}(E_z)$ created by the dipole in the $uz$ and $wz$ planes (Fig. 3E and F, left panels) with the analytically calculated $\mathrm{Re}(E_z)$ of the M0 and M1 modes propagating along [100] and [001] crystal axes, respectively (Fig. 3E and F, middle panels). The resemblance of the fields from the dipole with the analytically calculated electric field of the modes (see Section 5 in the Supporting Information) confirms the dominating role of the latter.

The identification of the dominating contributions of the polaritonic modes in a strongly anisotropic thin layer enables us to gain more insights into the structure of the electromagnetic fields created by localized sources, such as antennas, s-SNOM tips, defects or single quantum emitters. Particularly, the analysis of the field distribution across the layer can clarify the surface or volume character of the modes. The volume character of the mode has an important effect on both the losses (whether the fields are confined inside or outside of the layers) and its launching efficiency by localized sources. Additionally, the degree of confinement of the fields inside and outside of the layer affects the coupling of the modes to oscillations in molecular layers that can potentially be deposited on top of the biaxial crystal layers.[42,43] Thus, the modes shown in Figs. 3B, C, and E are classified as volume modes as they have a local maximum of their electric field amplitude at the center of the thin layer. In contrast, the mode shown in Fig. 3F has a signature of its surface character as the field decays towards the center of the layer. We note that, since the energy is proportional to the square of the amplitude of the electric field, it shows minima and maxima in the same locations as the absolute value of the electric field. To rigorously distinguish between surface-like and volume-like modes, we introduce the following criterion: we define volume modes as those presenting mostly real $z$ component of their wavevector inside the thin layer, while surface modes are those that mostly present imaginary $z$ component of their wavevector inside the thin layer. Following this definition, we introduce a measure, $\xi$, of the surface or volume nature of a mode, given by the expression $\xi = \tan^{-1}\frac{\mathrm{Im}\,k_z}{\mathrm{Re}\ _z}$, where $k_z$ is the out-of-plane component of the wavevector inside the biaxial thin layer, calculated with the Fresnel equation (see Section 1 in the Supporting Information). Accordingly, $\xi = 0$ and $\xi = {}^{\pi}/_{2}$ represent the limiting cases of volume and surface modes, respectively. The value of $\xi$ for the M0 mode as a function of frequency and direction of its wavevector is represented by a colorplot in Fig. 3G, where the white areas correspond to the absence of modes and the blue and red areas represent purely volume and surface modes, respectively. Analogously, Fig. 3H shows a colorplot of $\xi$ for the M1 mode. According to both the value of $\xi$ and the field distributions, the surface character of the fundamental mode M0 is clear in several frequency ranges, where $\varepsilon_z$ and at least one of the in-plane components of $\hat{\varepsilon}$ are negative (see details in Section 6 of the Supporting Information). In contrast, the M1 mode cannot be classified as a surface mode in any area of the in-plane-angle frequency space. This can be explained by the non-decaying out-of-plane field component inside the thin layer so that a larger portion of the electromagnetic field is confined inside its volume, which holds for all high-order modes.

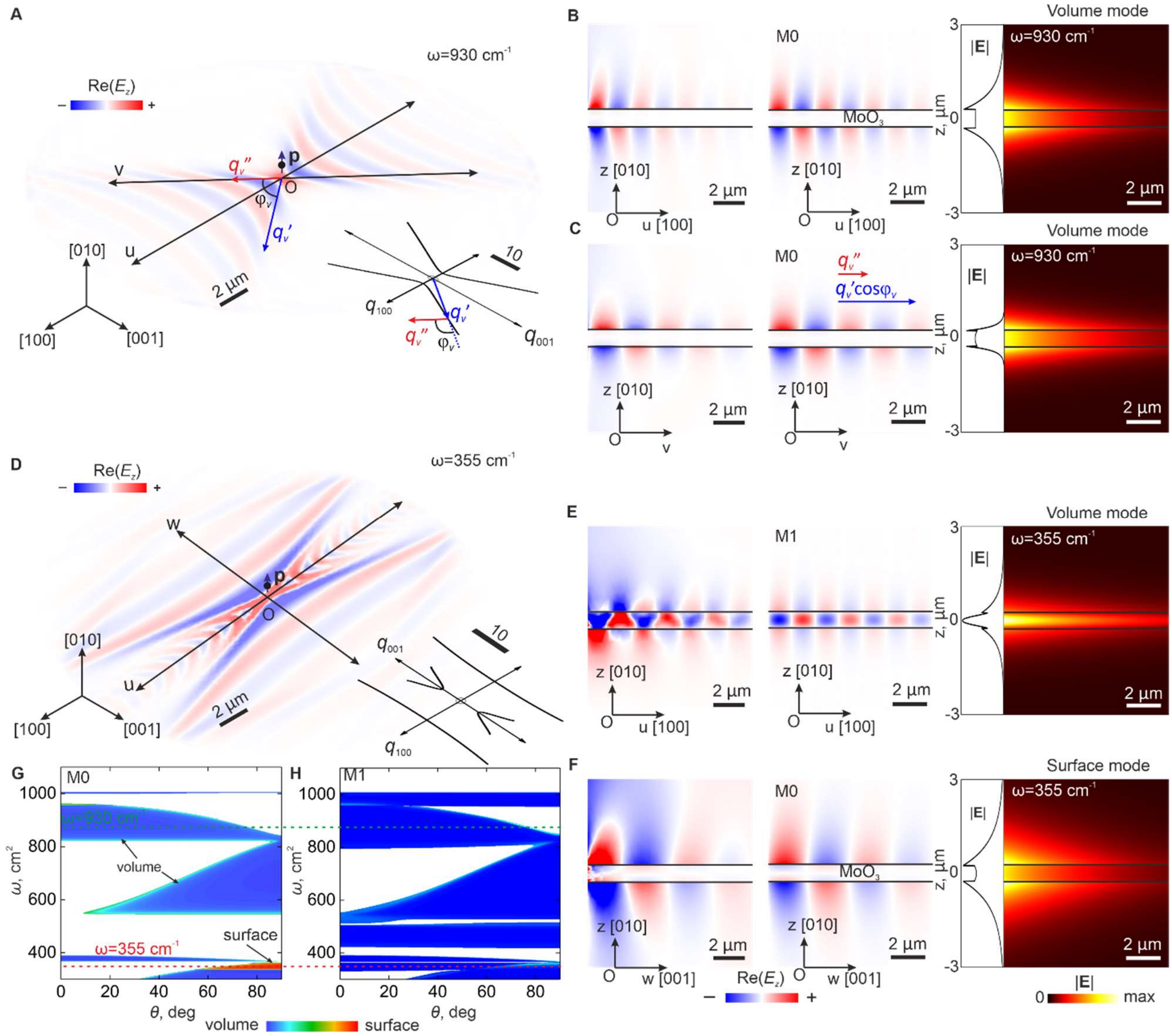

**Figure 3. Illustration of different hyperbolic modes and their contribution to the fields from a localized source. A)** FEM simulation of $\mathrm{Re}(E_z)$ created by a vertically oriented electric dipole (placed at a height of $500\ \mathrm{nm}$ above the layer). The field is calculated at $\omega = 930\ \mathrm{cm}^{-1}$ and extracted at a height of $50\ \mathrm{nm}$ above the layer surface. The insert shows the IFC for the same frequency with the wavevector and group velocity propagating in the *v*-direction. **B)** A vertical cross-section of the $\mathrm{Re}(E_z)$ obtained for an electric dipole (Left) and for the M0 mode (Right) along the *u*-direction (parallel to the α-$MoO_3$ [100] axis). The panel on the right shows the spatial distribution of $|\boldsymbol{E}|$ of the M0 mode in the $uz$ plane. **C)** Same as (**B**) for the *v*-direction. **D)** Same as (**A**) for $\omega = 355\ \mathrm{cm}^{-1}$. **E)** and **F)** show the results analogous to (**B**) and (**C**) along the *u*- and *w*-directions, respectively. $|\mathbf{E}|$ as a function of $z$ in the right panel in **F)** indicates the surface type of the M1 mode. **G)** and **H)** $\xi$ as a function of $\omega$ and direction of the real part of the in-plane wavevector with respect to the x-axis, $\theta$. Blue regions correspond to volume modes, $\mathrm{Re}(k_z) \gg \mathrm{Im}(k_z)$ and red regions to surface modes, $\mathrm{Re}(k_z) \ll \mathrm{Im}(k_z)$, respectively. In the white regions the M0 and M1 modes do not exist. The thickness of the free-standing layer is the same as in Fig. 1.

Up to now, our analysis of polaritons in biaxial materials has been exclusively focused on highly confined TM-polarized modes. Nonetheless, biaxial layers also support TE-polarized modes. Specifically, when these modes travel along the primary axes of the layer, they can be entirely TE-polarized. Otherwise, typically, the modes are partially TE- and TM-polarized. As equations (1,2) present the dispersion relation of purely TM modes with large momenta, to analytically calculate the dispersion relation of TE modes in biaxial layers, we have to use a different approximation. In general, the dispersion relation of polaritons in arbitrarily thick biaxial layers does not split into two independent equations separately describing the dispersion relations for the TE and TM modes.

However, since thin layers present a substantial interest for nanooptics, we can exploit this characteristic and approximate the layer by a 2D conducting sheet (thus avoiding the calculation of the fields inside the layer[3,29,32]), which allows separating the TE and TM modes. In this approximation, the dispersion relation for polaritons reads:

$$\left\{\alpha_x q_y^2 + \alpha_y q_x^2 + \frac{q_x^2+q_y^2}{2}(q_{1z}+q_{2z})\right\}\left\{\alpha_x q_x^2 + \alpha_y q_y^2 + \frac{q_x^2+q_y^2}{2}\left(\frac{\varepsilon_1}{q_{1z}}+\frac{\varepsilon_2}{q_{2z}}\right)\right\} = q_x^2 q_y^2\left(\alpha_x-\alpha_y\right)^2, \quad (3)$$

where $\alpha_j = \frac{k_0 d\varepsilon_j}{2i}$ $(j = x, y)$ is the normalized 2D conductivity along the $j$-direction, $q_{1z}$, and $q_{2z}$ are the normalized $z$-components of the wavevector in the superstrate, and the substrate, respectively. Equation (3) provides the dispersion of the fundamental (lowest-momentum) modes of both polarizations. In the particular case of modes propagating along the main axes $x$ and $y$, we have $q_y = 0$ and $q_x = 0$, respectively. Therefore, Eq. (3) splits into two equations for the TM and TE modes, given by zeroing the left and right brackets in the l.h.s., respectively. However, for an arbitrary propagation direction, the TE and TM modes are coupled. Nevertheless, in the approximation of thin layers, $d \ll \lambda_0$, the values of the effective conductivities are small, $|\alpha_x|, |\alpha_y| \ll 1$, so that the r.h.s. of Eq. (3) can be neglected, thereby enabling the separation of the TM and TE modes (see Section 7 in the Supporting Information for more details):

$$\begin{aligned}\text{TE:}\quad & \alpha_x q_y^2 + \alpha_y q_x^2 + \frac{q_x^2+q_y^2}{2}(q_{1z}+q_{2z}) = 0,\\ \text{TM:}\quad & \alpha_x q_x^2 + \alpha_y q_y^2 + \frac{q_x^2+q_y^2}{2}\left(\frac{\varepsilon_1}{q_{1z}}+\frac{\varepsilon_2}{q_{2z}}\right) = 0.\end{aligned} \quad (4)$$

For a free-standing biaxial layer, i.e., when $\varepsilon_1 = \varepsilon_2 = 1$, we have $q_{1z} = q_{2z} = \sqrt{1-q^2}$, and the dispersion relation of the TE modes greatly simplifies:

$$q = \sqrt{1 + \frac{k_0^2 d^2}{4}\left(\varepsilon_y \cos^2\alpha + \varepsilon_x \sin^2\alpha\right)^2}. \quad (5)$$

Figure 4A represents the dispersion of the TE modes propagating along the main in-plane crystal axes of an α-$MoO_3$ layer of thickness $d = 500$ nm calculated by transfer matrix methods[41] (colorplot showing the maxima of the imaginary part of the Fresnel reflection coefficient), and using our analytical Eq. (5) (solid curves). A very good agreement is obtained. Analogously to the TM modes, the TE modes can either propagate along any in-plane direction or within angular sectors in elliptic or hyperbolic range, respectively. Note that these angular sectors are centered along the same crystal axes as the sectors of propagation of the TM modes at the same frequency, which can be verified by comparing Fig. 1C and Fig. 4A (see Section 7 in the Supporting Information for more details). Additionally, according to their dispersion, the TE modes have a much longer wavelength and weaker confinement than the TM modes. For example, at $\omega = 670\ \mathrm{cm}^{-1}$, $\lambda_{p\mathrm{TE}} = 10\ \mu\mathrm{m}$, while for the TM modes $\lambda_{p\mathrm{TM}} = 3.5\ \mu\mathrm{m}$. On the other hand, the TE modes have a larger propagation length in units of wavelength than the TM modes. For example, at $\omega = 670\ \mathrm{cm}^{-1}$, $L_{p\mathrm{TE}}/\lambda_{p\mathrm{TE}} = 8$, while $L_{p\mathrm{TM}}/\lambda_{p\mathrm{TM}} = 2.5$ for the TM modes. This implies that the TE modes have a better figure of merit (FOM), which can be relevant for potential applications requiring the propagation of polaritons across a device. To further explore the propagation properties of the TE modes, we perform a quasi-eigenmode mode analysis (full-wave numerical simulation) and plot the electric field distribution of the propagating eigenmodes for two frequencies within the elliptic ($\omega = 490\ \mathrm{cm}^{-1}$) and hyperbolic ($\omega = 670\ \mathrm{cm}^{-1}$) spectral ranges (Fig. 4B). The spatial field distribution confirms that the TE modes present a longer wavelength and weaker confinement than the TM modes. Longer wavelength of the mode implies larger sizes of potential polaritonic resonators which can be used in sensing or probing light-matter interactions.[44]

To investigate the propagation of the TE modes along an arbitrary direction with respect to the main crystal axes, we plot their IFCR and IFCI for two representative frequencies in both the elliptic, $\omega = 490\ \mathrm{cm}^{-1}$, and the hyperbolic, $\omega = 670\ \mathrm{cm}^{-1}$, frequency ranges. To that end, we use Eq. (3) and follow the same minimization procedure as the one used before to calculate the IFCs of the TM modes. Figure 4C shows the IFCR and IFCI of the TE modes at $\omega = 490\ \mathrm{cm}^{-1}$, where the modes have an elliptic IFC. We observe that the real and the imaginary parts of the wavevector of the TE modes are not parallel, analogously to the TM modes. In addition, while the IFCR and IFCI of the TE modes are ellipse-like, they exhibit a remarkably elongated shape along one of the axes, in contrast to the IFCR and IFCI of the TM modes (compare Fig. 4C and Fig. 2B). Such misalignment of the IFCs makes the FOM of the TE modes to be significantly dependent upon the propagation direction. An even more critical difference between the TE and TM modes appears in the hyperbolic range: while the wavevector of the TM modes increases with the angle that it forms with respect to the transverse axis of the hyperbola (Fig. 2A), the wavevector of the TE modes decreases with the angle until it reaches the light cone (Fig. 4D). This behavior of the IFC is similar to that of “leaky” polaritons – surface modes that radiate into the substrate or superstrate –,[13] however, the TE modes have a different nature, since they are completely confined within an anisotropic layer. Additionally, one can clearly recognize flat regions of the IFCR of the TE modes in the hyperbolic range. As the group velocity is aligned perpendicularly to the IFCR within these flat regions, TE-polarized polaritons can exhibit canalization phenomena. To explore such unique possibility, we perform full-wave numerical simulations. Note that, in contrast to the TM modes, which can be excited by electric dipoles, the TE modes are much more efficiently launched by magnetic dipoles since they provide a significant vertical component of the magnetic field. Therefore, we simulate the magnetic field created by a vertically oriented magnetic dipole source placed near the surface of the $\alpha$-$MoO_3$ thin layer. To better visualize the TE modes, we suppress the excitation of the TM modes by placing the magnetic dipole sufficiently far from the surface, e.g., at a distance of $h = 10\ \mu\mathrm{m}$, which is much larger than the wavelength of any TM mode launched by the dipole. The simulated vertical component of the magnetic field is represented in Fig. 4E, while Fig. 4F shows its amplitude. The latter clearly shows that a significant portion of the electromagnetic energy propagates within four narrow beams, unambiguously demonstrating that polariton canalization can be achieved for the TE polaritonic modes in thin biaxial layers.

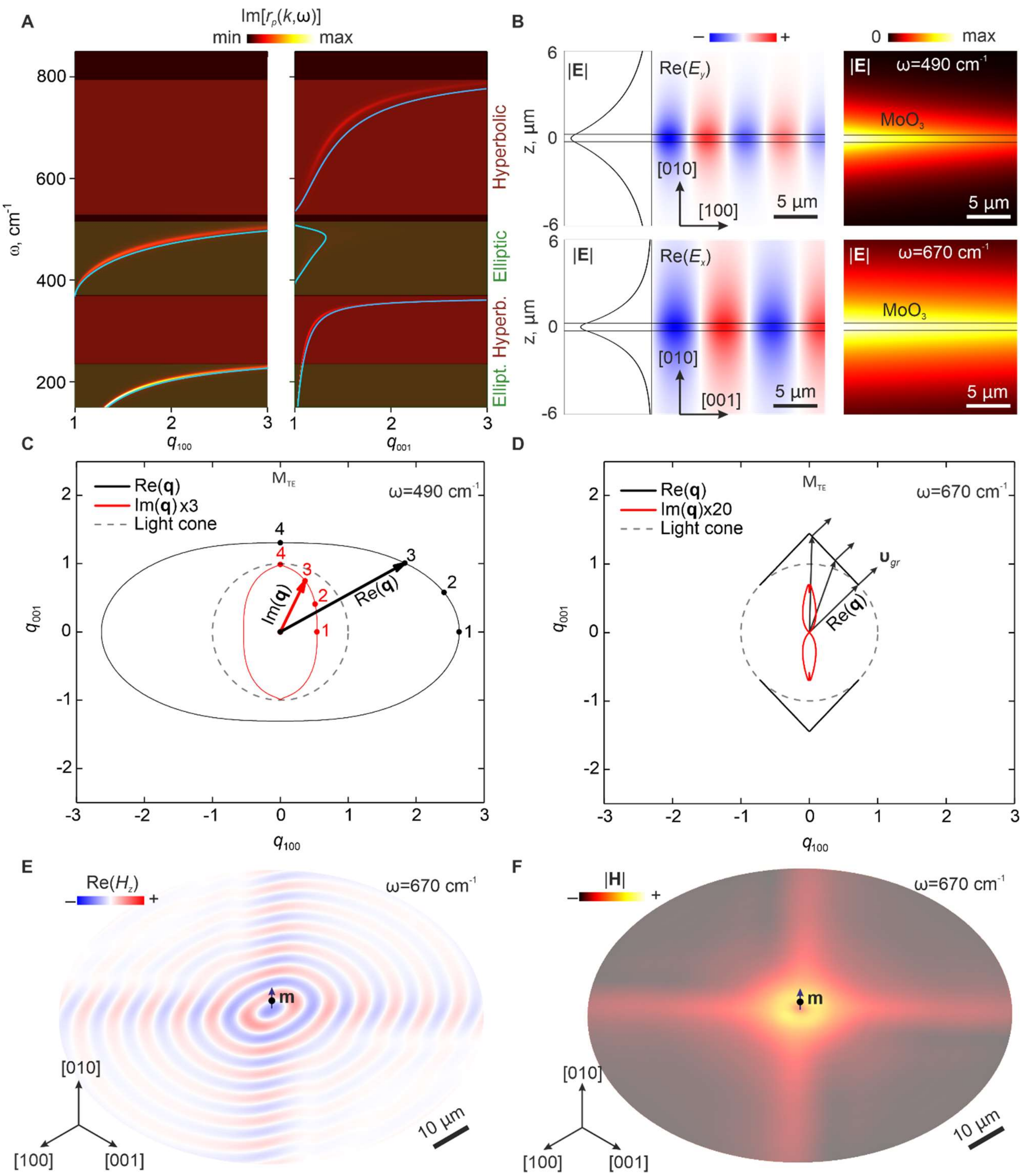

**Figure 4. TE modes in a free-standing biaxial layer. A)** Dispersion relations of TE modes propagating along the main [100] and [001] crystal directions. Frequency is shown as a function of the normalized wavevector $q = \frac{k}{k_0}$. The colormaps represent the imaginary part of the Fresnel reflection coefficient obtained by transfer matrix calculations, while the blue solid curves are the result of analytical calculations using Eq. (5). **B)** Electric field distributions of the TE modes (the electric field has only the field component perpendicular to the wavevector) propagating along the main crystal directions calculated for $\omega = 490\ \mathrm{cm}^{-1}$ and $\omega = 670\ \mathrm{cm}^{-1}$. **C)** IFC for the TE mode in the elliptic range ($\omega = 490\ \mathrm{cm}^{-1}$). Analogously to the TM modes, the real and imaginary parts of the wavevector are not parallel to each other. **D)** IFC for the TE mode in the hyperbolic range ($\omega = 670\ \mathrm{cm}^{-1}$). The flat regions in the IFC yield the canalization of polaritons along two specific directions. **E)** and **F)** Full-wave numerical simulations representing the vertical component of the magnetic field, $H_z$, inside the layer (**E**) and the absolute value of the magnetic field, $|\mathbf{H}|$, inside the layer (**F**), respectively. A vertically oriented magnetic dipole is placed at the height $h = 10\ \mu\mathrm{m}$ above the layer surface and is used as an excitation source ($\omega = 670\ \mathrm{cm}^{-1}$). The thickness of the free-standing layer is the same as in Figs. 1,2.

## Conclusions

We have introduced a paradigm shift in the analysis of polaritons in thin layers of a strongly anisotropic medium (using the van der Waals crystal α-$MoO_3$ as an example), highlighting the crucial role of the optical losses. Namely, we have shown that the assumption of collinearity of the real and imaginary parts of the wavevector, which were used hitherto, is not general and could lead to imprecise results and wrong interpretation of near-field optical experiments. In particular, we have demonstrated that each fixed direction of $\mathbf{k}'$ allows different directions of $\mathbf{k}''$, so that the latter has to be determined with an additional assumption: minimizing the losses. Basing on the latter, we have developed a novel analytical tool allowing one to visualize polaritonic IFCs in a multidimensional space of complex wavevectors. Visualization of IFCs enables providing insights into the peculiarities of polaritons, particularly in the hyperbolic frequency ranges, where $\mathbf{k}'$ is close to the asymptotes of the IFC. These insights are crucial for studies of refraction and reflection and other fundamental optical phenomena. We have also suggested a rigorous criterion to distinguish between the surface and volume nature of highly confined electromagnetic waves in strongly anisotropic layers. Finally, we have identified the TE polaritonic modes, up to now lacking in the literature, opening a new direction for research on electromagnetic modes in biaxial crystal layers. One of the crucial results of our analysis is the prediction of the TE mode canalization in one of the hyperbolic spectral ranges, due to their strongly anisotropic and flat IFCs at some frequencies. We believe that our results expand the understanding of electromagnetic waves in highly anisotropic thin layers, offering a deeper understanding of the key characteristics of anisotropic polaritons and thereby opening the door to new possibilities for their practical applications.

## Acknowledgements

The project that gave rise to these results received the support of a fellowship from “la Caixa” Foundation (ID 100010434). The fellowship code is LCF/BQ/DI21/11860026. G. Á.-P. Acknowledges support from the Severo Ochoa program of the Government of the Principality of Asturias (grant number PA-20-PF-BP19-053). P.A.-G. acknowledges support from the European Research Council under Consolidator grant no. 101044461, TWISTOPTICS and the Spanish Ministry of Science and Innovation (grant numbers PID2019-111156GB-I00 and PID2020-115221GB-C42).

## Supporting Information

The Supporting Information contains the derivation of the dispersion relation in high-q approximation considering the TM modes only; the description of the method of the calculation of the isofrequency curves by minimizing the losses; the derivation of the expressions for the directions of the group velocity and the imaginary part of the wavevector; the comparison of the IFC obtained with the different approaches; the analytical expression for the field distribution; conditions for the surface or volume type of the mode; the derivation of the dispersion relation of the TE modes in the ultrathin-slab limit.

The Supporting Information is available free of charge at http://pubs.acs.org

# Supplementary Material for

# The fundamentals of lossy anisotropic polaritons in biaxial crystal slabs

Kirill V. Voronin[1], Gonzalo Álvarez-Pérez[2,3], Christian Lanza[2], Pablo Alonso-González[2,3,†], Alexey Y. Nikitin[1,4,‡]

[1]Donostia International Physics Center (DIPC), Donostia-San Sebastián 20018, Spain.
[2]Department of Physics, University of Oviedo, Oviedo 33006, Spain.
[3]Center of Research on Nanomaterials and Nanotechnology, CINN (CSIC-Universidad de Oviedo), El Entrego 33940, Spain.
[4]IKERBASQUE, Basque Foundation for Science, Bilbao 48013, Spain.

† Corresponding author: pabloalonso@uniovi.es
‡ Corresponding author: alexey@dipc.org

## 1. The derivation of Eq. (1) considering the TM modes only

Generally, TE and TM modes in a biaxial slab are coupled, and the dispersion relation of the in-plane modes include both "o"- and "e"-modes in the biaxial slab and s- and p-waves in free space.[1] In ref. [1], Eq. (1) of the main text was derived under the assumption that the polaritonic wavevector $k$ is much larger than the free-space light wavevector $k_0 = \frac{\omega}{c}$, $k \gg k_0$. From here on, we will refer to this limit as the high-$k$ approximation. Under such approximation, the general dispersion equation greatly simplifies resulting in Eq. (1) of the main text, however, the polarization of the obtained mode remains unclear. Here we perform an alternative derivation of Eq. (1) of the main text demonstrating that only the TM modes contribute to the dispersion relation under the high-$k$ approximation. To do this, we start by simplifying the basis vectors for both polarizations "o" and "e" in biaxial media, which are given by[1]:

$$|o\rangle = \frac{1}{q}\begin{pmatrix} -q_y(1-\Delta_1\Delta_z) \\ q_x \\ -q_x q_y q_{oz}\Delta_1 \end{pmatrix} e^{ik_x x + ik_y y}, \qquad |e\rangle = \frac{1}{q}\begin{pmatrix} q_x \frac{\Delta_2 - q_y^2}{\Delta_x^e} \\ q_y \\ \frac{\Delta_2}{-q_{ez}} \end{pmatrix} e^{ik_x x + ik_y y}, \tag{S1}$$

where $q_i$ $(i = x, y, z)$ are the normalized wavevector projections, $q_i = \frac{k_i}{k_0}$, $q = \sqrt{q_x^2 + q_y^2}$; $\Delta_i^j = \varepsilon_i - q_x^2 - q_y^2 - q_{jz}^2 + q_i^2$ $(i = x, y, j = o, e)$, $\Delta_z = \varepsilon_z - q_x^2 - q_y^2$, and $\Delta_1$ and $\Delta_2$ are given by:

$$\Delta_1 = \frac{\Delta_x^o - q_x^2}{\Delta_x^o \Delta_z - q_x^2 q_{oz}^2}, \qquad \Delta_2 = \frac{\Delta_x^e \Delta_y^e - q_x^2 q_y^2}{\Delta_x^e - q_x^2}. \tag{S2}$$

Under the high-$k$ approximation, the Fresnel equation can be significantly simplified, and $q_{oz}$ and $q_{ez}$ can be expressed in terms of $q_x$ and $q_y$ as follows:

$$q_{oz}^2 = -q_x^2 - q_y^2, \qquad q_{ez}^2 = -\frac{\varepsilon_x}{\varepsilon_z} q_x^2 - \frac{\varepsilon_y}{\varepsilon_z} q_y^2. \tag{S3}$$

In this case, we can write $\Delta_1 = \frac{1}{q_{oz}^2}\frac{\varepsilon_x - \varepsilon_y}{\varepsilon_x - \varepsilon_z}$ and $\Delta_2 = -q_{ez}^2$. The polarization basis vectors can then be written as follows:

$$|o\rangle = \frac{1}{q}\begin{pmatrix} -q_y(\varepsilon_y - \varepsilon_z) \\ q_x(\varepsilon_x - \varepsilon_z) \\ -\frac{q_x q_y}{q_{oz}}(\varepsilon_x - \varepsilon_y) \end{pmatrix} e^{ik_x x + ik_y y}, \qquad |e\rangle = \frac{1}{q}\begin{pmatrix} q_x \\ q_y \\ q_{ez} \end{pmatrix} e^{ik_x x + ik_y y}, \tag{S4}$$

Thus, within the high-$k$ approximation, the "e"-mode of the biaxial medium becomes longitudinal, which means that the boundary condition could be satisfied with only "e"-polarized waves inside the slab and the p-polarized wave outside. In other words, longitudinal and transverse fields become independent and could be considered separately. Considering only the TM waves, we can write the fields in all three media:

$$\begin{gathered} \vec{E}_1 = a_1 \begin{pmatrix} q_x \\ q_y \\ iq \end{pmatrix} e^{i(k_x x + k_y y)} e^{-qk_0 z}; \quad z \geq d \\ \vec{E}_2 = a_2 \begin{pmatrix} q_x \\ q_y \\ q_{ez} \end{pmatrix} e^{i(k_x x + k_y y)} e^{iq_{ez}k_0 z} + b_2 \begin{pmatrix} q_x \\ q_y \\ -q_{ez} \end{pmatrix} e^{i(k_x x + k_y y)} e^{-iq_{ez}k_0 z}; \quad d > z \geq 0 \\ \vec{E}_3 = b_3 \begin{pmatrix} q_x \\ q_y \\ -iq \end{pmatrix} e^{i(k_x x + k_y y)} e^{qk_0 z}; \quad z < 0, \end{gathered} \tag{S5}$$

where $\vec{E}_1$, $\vec{E}_2$, and $\vec{E}_3$ are fields in the superstrate, substrate, and slab respectively; $z = 0$ is the coordinate of the bottom surface of the slab; here we also take into account that, in the high-$k$ approximation, the p-polarization basis vector is given by $|p\rangle = (q_x \quad q_y \quad iq)^{\mathrm{T}} e^{i(k_x x + k_y y)}$.

Imposing the boundary conditions for the in-plane components, $\vec{E}_{1||}(z = d) = \vec{E}_{2||}(z = d)$, $\vec{E}_{2||}(z = 0) = \vec{E}_{3||}(z = 0)$ and for the out-of-plane components, $\vec{D}_{1\perp}(z = d) = \vec{D}_{2\perp}(z = d)$, $\vec{D}_{2\perp}(z = 0) = \vec{D}_{3\perp}(z = 0)$, we obtain the following system of equations:

$$\begin{gathered} a_1 e^{-qk_0 d} = a_2 e^{iq_{ez}k_0 d} + b_2 e^{-iq_{ez}k_0 d} \\ a_1 i\varepsilon_1 q e^{-qk_0 d} = a_2 \varepsilon_z q_{ez} e^{iq_{ez}k_0 d} - b_2 \varepsilon_z q_{ez} e^{-iq_{ez}k_0 d} \\ a_2 + b_2 = b_3 \\ a_2 \varepsilon_z q_{ez} - b_2 \varepsilon_z q_{ez} = -b_3 i\varepsilon_2 q. \end{gathered} \tag{S6}$$

Or in the matrix form:

$$\begin{pmatrix} e^{-qk_0 d} & -e^{iq_{ez}k_0 d} & -e^{-iq_{ez}k_0 d} & 0 \\ i\varepsilon_1 q e^{-qk_0 d} & -\varepsilon_z q_{ez} e^{iq_{ez}k_0 d} & \varepsilon_z q_{ez} e^{-iq_{ez}k_0 d} & 0 \\ 0 & 1 & 1 & -1 \\ 0 & \varepsilon_z q_{ez} & -\varepsilon_z q_{ez} & i\varepsilon_2 q \end{pmatrix} \begin{pmatrix} a_1 \\ a_2 \\ b_2 \\ b_3 \end{pmatrix} = 0 \tag{S7}$$

System (S7) has a solution if the determinant of the matrix of this system equals zero. Simplifying the determinant and equating it to zero, we obtain the dispersion relation of the mode:

$$\tan q_{ez} k_0 d + \frac{(\varepsilon_1 + \varepsilon_2)\varepsilon_z q q_{ez}}{\varepsilon_1 \varepsilon_2 q^2 + \varepsilon_z^2 q_{ez}^2} = 0. \tag{S8}$$

Which brings us to the explicit expression for k:

$$k = \frac{\rho}{d}\left[\operatorname{atan}\left(\frac{\rho\varepsilon_1}{\varepsilon_z}\right) + \operatorname{atan}\left(\frac{\rho\varepsilon_2}{\varepsilon_z}\right) + \pi l\right], \qquad l = 0,1, \dots \tag{S9}$$

where $\rho = i\sqrt{\frac{\varepsilon_z}{\varepsilon_x \cos^2\alpha + \varepsilon_y \sin^2\alpha}}$.

For the case of high dielectric permittivity of the slab, $|\varepsilon_x| \sim |\varepsilon_y| \sim |\varepsilon_z| \gg 1$, we can write $\mathrm{atan}\left(\frac{\rho\varepsilon_1}{\varepsilon_z}\right) \approx \frac{\rho\varepsilon_1}{\varepsilon_z}$ and $\mathrm{atan}\left(\frac{\rho\varepsilon_2}{\varepsilon_z}\right) \approx \frac{\rho\varepsilon_2}{\varepsilon_z}$, therefore, Eq. (S9) is significantly simplified:

$$k = -\frac{\varepsilon_1+\varepsilon_2}{d(\varepsilon_x \cos^2\alpha + \varepsilon_y \sin^2\alpha)} \tag{S10}$$

### 2. Calculation of the isofrequency curves by minimizing the losses

In order to calculate the solution of the generalized dispersion, Eq. (2) of the main text, corresponding to the minimum of losses, we have developed a simplifying procedure consisting of six steps (i-vi).

i. First, we reduce the space dimension by assuming that the frequency is real-valued: $\omega = \omega'$. Such an assumption is reasonable for the analysis of the propagation of polaritonic modes excited by continuous monochromatic radiation.

ii. Second, we take a cross-section of the hypersurface by a plane of a constant $\omega$, thus slicing an isofrequency hypersurface in four-dimensional space, $(k_x', k_x'', k_y', k_y'')$. To proceed with the next simplification step, we notice that, in practice, the excited polaritonic modes do not contribute equally to the spatial field distributions created by localized sources in realistic structures (such as resonant metallic antennas[2] or s-SNOM tips[3]) because of the different decay rate, which is given by the imaginary part of the wavevector. Particularly, sufficiently far from a source, the field is mostly composed of modes having larger propagation lengths. Consequently, we can restrict ourselves to considering only modes with lower losses, thus significantly simplifying the isofrequency hypersurface. In order to neglect the modes with large losses, we perform a minimization of the imaginary part of the wavevectors for a given propagation direction.

iii. To that end we change the coordinate system from $(k_x', k_x'', k_y', k_y'')$ to $(k', k'', \alpha_1, \alpha_2)$, where $k'$ and $k''$ are the absolute values of the real and the imaginary parts of the wavevector, and $\alpha_1$ and $\alpha_2$ are the angles that they form with the $x$-axis, respectively. The transformation between these coordinate systems is given by the following:

$$\begin{aligned} k_x' &= k' \cos\alpha_1\,, \\ k_x'' &= k'' \cos\alpha_2\,, \\ k_y' &= k' \sin\alpha_1\,, \\ k_y'' &= k'' \sin\alpha_2\,, \end{aligned} \tag{S11}$$

that is, $k', k'' \in [0, \infty)$ and $\alpha_1, \alpha_2 \in [0, 2\pi)$.

iv. We look for points on the isofrequency hypersurface that yield polaritons with the smallest imaginary parts of their wavevectors, $k''$, for each direction of the real part, given by $\alpha_1$. Technically, returning back to Eq. (2), this means that we look for its solutions with the minimal $k''$, for a fixed frequency and a fixed $\alpha_1$. We vary $\alpha_2$ from 0 to $2\pi$. For the given $\alpha_1$ and $\alpha_2$, Eq. (2) allows us to find the real and imaginary parts of the wavevector, $k_s'(\alpha_1,\alpha_2)$, $k_s''(\alpha_1,\alpha_2)$, satisfying two equations, $\mathrm{Re}[f(k_x,k_y,\omega)] = 0$ and $\mathrm{Im}[f(k_x,k_y,\omega)] = 0$, for two real-valued unknowns, $k', k''$. Then, we extract the minimum value from the $k_s''(\alpha_1,\alpha_2)$, and denote it as $k_c''(\alpha_1)$. The latter is given by $k_c''(\alpha_1) = \min_{\alpha_2\in[0,2\pi)} k_s''(\alpha_1,\alpha_2)$. The corresponding value of the angle $\alpha_2$ providing the minimal value of $k_s''(\alpha_1,\alpha_2)$ we denote as $\alpha_{2\mathrm{c}}(\alpha_1)$. More explicitly, the definition for the $k_c''$ reads as $k_c''(\alpha_1) = \min_{\alpha_2 \in [0,2\pi)} k_s''(\alpha_1, \alpha_2) = k_s''(\alpha_1, \alpha_{2c}(\alpha_1))$. Finally, we find the absolute value of the real part of the wavevector as $k_c'(\alpha_1) = k_s'(\alpha_1, \alpha_{2c}(\alpha_1))$. Thus, for each $\alpha_1$ for which Eq. (2) has a solution, we calculate $\alpha_{2c}(\alpha_1), k_c''(\alpha_1)$ and $k_c'(\alpha_1)$. A geometrical interpretation of step (iv) can be seen as reducing the isofrequency hypersurface in four-dimensional space $(k', k'', \alpha_1, \alpha_2)$ to an isofrequency curve (IFC). The latter can be parametrically defined as $(k_c'(\alpha_1), k_c''(\alpha_1), \alpha_1, \alpha_{2c}(\alpha_1))$, $\alpha_1 \in [0,2\pi)$.

v. We return to the original coordinates, ($k_x'$, $k_x''$, $k_y'$, $k_y''$). In the original coordinates the IFC obtained in the previous step can also be represented parametrically, with the parameter $\alpha_1$ ($\alpha_1 \in [0,2\pi)$), as follows: ($k_{xc}'(\alpha_1)$, $k_{xc}''(\alpha_1)$, $k_{yc}'(\alpha_1)$, $k_{yc}''(\alpha_1)$).

vi. Finally, for the visualization of the IFC, we plot its projections in the planes ($k_x'$, $k_y'$) and ($k_x''$, $k_y''$). We denote these projections as real (IFCR) and imaginary (IFCI) IFCs, respectively. Note, that for the IFCR, $\alpha_1$ is the polar angle, $\tan \alpha_1 = k_y'/k_x'$, whereas the polar angle of IFCI, $\alpha_2$, can depend on $\alpha_1$ in a non-trivial way. Importantly, if losses are not high, $\mathrm{Im}(\varepsilon_i) \ll |\varepsilon_i|$ $(i = x, y, z)$, then under the condition of minimal $k''$, the vector $\mathbf{k}''$ is parallel to the group velocity vector, see Supporting Information, section 4 for the details. This result means that polaritons launched by a dipolar source have both the group velocity and imaginary part of the wavevector aligned with the radius vector. The latter starts at the position of the source and ends at the observation point. In other words, the field decays along the direction of propagation.

### 3. The directions of the group velocity and the imaginary part of the wavevector

Let us consider the dispersion equation in a general form $F\left(k_x, k_y, \omega, \varepsilon_i(\omega)\right) = 0$, where $\varepsilon_i(\omega)$ is the set of all of the dielectric permittivities in the system. For a monochromatic process, it defines the relation between the in-plane components of the wavevector, $k_x$ and $k_y$. For lossless media, characterized by real $\varepsilon_i(\omega)$, this expression directly gives the IFC. To find the group velocity for each allowed wavevector we follow the definition $\boldsymbol{\upsilon}_{gr} = d\omega/d\mathbf{k}$ and calculate the gradient of $F\left(k_x, k_y, \omega, \varepsilon_i(\omega)\right)$, we obtain the following

$$\boldsymbol{\upsilon}_{gr} = \frac{d\omega}{d\mathbf{k}} = -\frac{\partial F/\partial \mathbf{k}}{\partial F/\partial \omega} \tag{S12}$$

This means that the group velocity is perpendicular to the IFC given by the equation $F\left(k_x, k_y, \omega, \varepsilon_i(\omega)\right) = 0$.

Now we allow $\varepsilon_i$ to have the imaginary part, introducing the losses into the system, but we assume that the losses are low, $\mathrm{Im}[\varepsilon_i] \ll \mathrm{Re}[\varepsilon_i]$. As a result, $k_x$ and $k_y$ also acquire the imaginary parts; therefore, we get the system of two equations

$$\begin{cases} \mathrm{Re}\left[F\left(k_x' + ik_x'', k_y' + ik_y'', \omega, \varepsilon_i' + i\varepsilon_i''\right)\right] = 0 \\ \mathrm{Im}\left[F\left(k_x' + ik_x'', k_y' + ik_y'', \omega, \varepsilon_i' + i\varepsilon_i''\right)\right] = 0 \end{cases} \tag{S13}$$

with four variables, $k_x'$, $k_x''$, $k_y'$, $k_y''$. Considering imaginary parts as small terms and performing the expansion of $F(k_x, k_y, \omega, \varepsilon_i)$ up to the first order we obtain the following:

$$F(k_x, k_y, \omega, \varepsilon_i) = F(k_x', k_y', \omega, \varepsilon_i') + i\mathbf{k}'' \frac{\partial F(k_x', k_y', \omega, \varepsilon_i')}{\partial \mathbf{k}'} + i\varepsilon_i''(\omega) \frac{\partial F(k_x', k_y', \omega, \varepsilon_i')}{\partial \varepsilon_i'} \tag{S14}$$

Here we assume summation over the $i$-index. Consequently, the first equation of the system (S13) brings us to the IFC for the lossless case, $F\left(k'_x, k'_y, \omega, \varepsilon'_i\right) = 0$, whereas the second one defines the imaginary part of the wavevector, more precisely, its projection on the group velocity. Taking into account Eq. (S14) we obtain

$$\mathbf{k}''\boldsymbol{\upsilon}_{gr}\frac{\partial F\left(k'_x,k'_y,\omega,\varepsilon'_i(\omega)\right)}{\partial\omega} = \varepsilon''_i\frac{\partial F\left(k'_x,k'_y,\omega,\varepsilon'_i\right)}{\partial\varepsilon'_i} \tag{S15}$$

Thus, minimization of $|\mathbf{k}''|$ requires $\mathbf{k}''$ to be parallel to the group velocity or to be perpendicular to the real part of the IFC. Interestingly, if $F\left(k'_x, k'_y, \omega, \varepsilon'_i\right)$ does not depend on $\omega$ explicitly, for example, in the high-$k$ approximation, we can write

$$\frac{\partial f\left(k'_x,k'_y,\varepsilon'_i(\omega)\right)}{\partial\omega} = \frac{\partial f\left(k'_x,k'_y,\varepsilon'_i\right)}{\partial\varepsilon'_i}\frac{d\varepsilon'_i}{d\omega} \tag{S16}$$

As a result, Eq. (S15) could be simplified as follows

$$\mathbf{k}''\boldsymbol{\upsilon}_{gr}\frac{d\varepsilon'_i}{d\omega} = \varepsilon''_i \tag{S17}$$

Therefore, according to Eq. (S17) the expression for the minimal $\mathbf{k}''$ reads

$$\mathbf{k}'' = \frac{\varepsilon''_i\boldsymbol{\upsilon}_{gr}}{\left|\boldsymbol{\upsilon}_{gr}\right|^2\frac{d\varepsilon'_i}{d\omega}} \tag{S18}$$

Taking into account that $\frac{d\varepsilon'_i}{d\omega} > 0$ everywhere except the narrow high-loss ranges, which are out of the scope of our work, minimization of the imaginary part of the wavevector requires $\mathbf{k}''$ to be co-directional to $\boldsymbol{\upsilon}_{gr}$. On the other hand, if $\mathbf{k}'' \nparallel \boldsymbol{\upsilon}_{gr}$, the expression for $\mathbf{k}''$ reads

$$k'' = \frac{\varepsilon''_i}{\upsilon_{gr}\cos\beta\frac{d\varepsilon'_i}{d\omega}} \tag{S19}$$

where $\beta$ is the angle between $\mathbf{k}''$ and $\boldsymbol{\upsilon}_{gr}$. In particular, if $\beta \to \frac{\pi}{2}$, $k'' \to \infty$, and the assumption of small losses breaks. For example, within the assumption $\mathbf{k}' \parallel \mathbf{k}''$, it leads to the appearance of the bending of the IFCR back to the origin, which has been theoretically reported but not observed in experiments.[4]

### 4. Comparison of the IFC obtained with the different approaches

In this section, we show the difference between the IFCs obtained by the minimization of the losses and IFCs obtained by the standard approach, under which $\mathbf{q}'$ and $\mathbf{q}''$ are assumed to be collinear. To do so, we plot the IFCR and the IFCI using these two different approaches for the frequencies used in Fig. 2, that is, $\omega = 955\ \mathrm{cm}^{-1}$ and $\omega = 980\ \mathrm{cm}^{-1}$. In the hyperbolic range, $\omega = 955\ \mathrm{cm}^{-1}$, the IFCs are touching in the vertex of the hyperbolic curve and then diverging near the asymptotes, see Figs. S1 A, C. For $\omega = 980\ \mathrm{cm}^{-1}$, the IFCRs are extremely close to each other, however, the IFCIs deviate remarkably. This could be explained by the smallness of $\mathbf{q}''$ with respect to $\mathbf{q}'$, indeed, since $|\mathbf{q}''| \ll |\mathbf{q}'|$, a relatively large variation of $\mathbf{q}''$ causes a relatively tiny variation of $\mathbf{q}'$ (Fig. S1B). Noticeably, at first glance, the $\mathbf{q}''$ obtained by the minimization of the losses for $\omega = 980\ \mathrm{cm}^{-1}$ seems to be larger than the $\mathbf{q}''$ obtained with the assumption of the collinearity of $\mathbf{q}'$ and $\mathbf{q}''$. However, in Fig. S1D we plot $\mathbf{q}''$ for a mode with an arbitrary direction of $\mathbf{q}'$, demonstrating that $\mathbf{q}''$ is smaller when obtained by minimizing the losses.

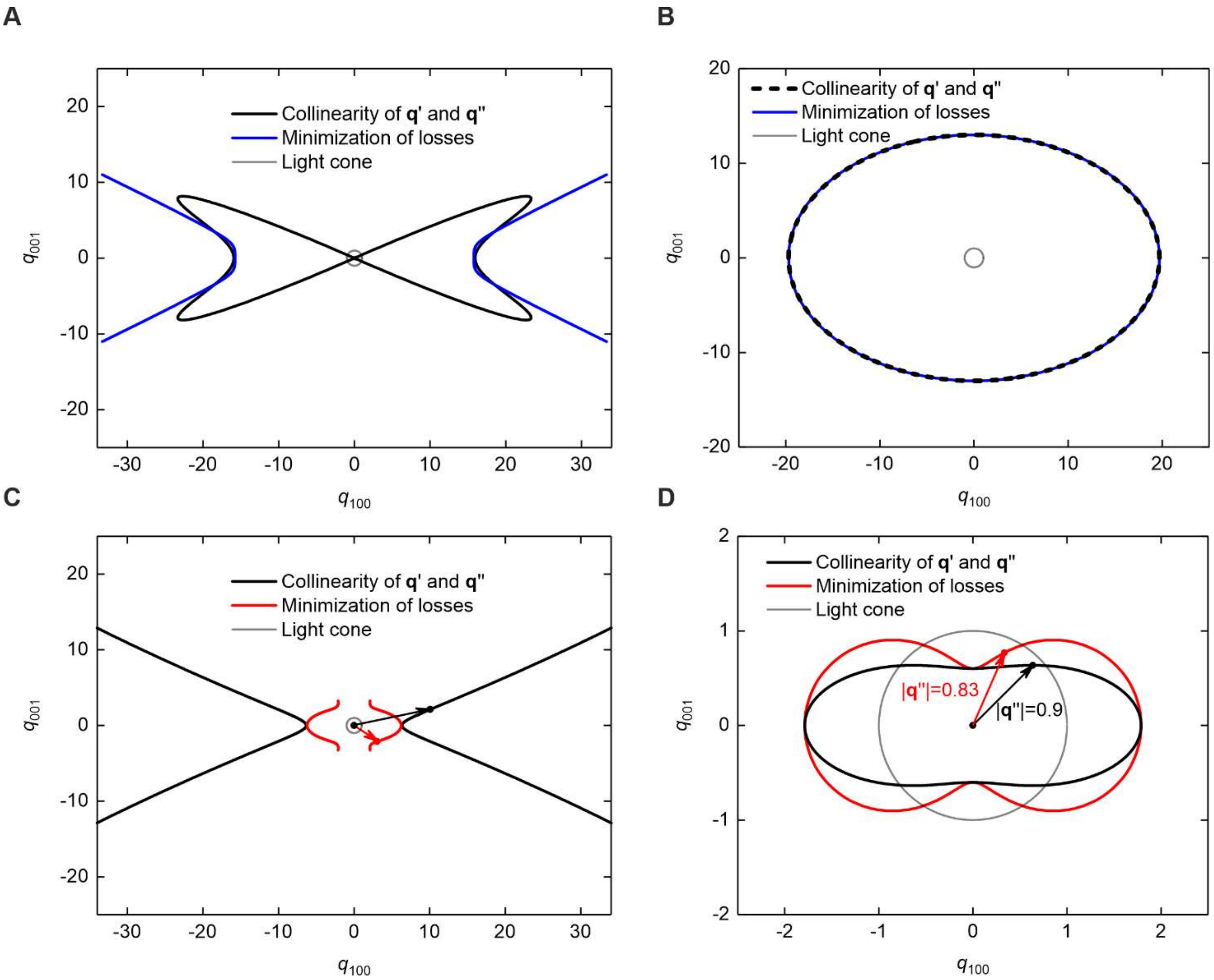

**Figure S1.** Comparison of the IFCR (**A** and **B**) and IFCI (**C** and **D**) obtained by the minimization of losses with the IFC obtained assuming $\mathbf{q}'$ and $\mathbf{q}''$ to be parallel, for two representing frequencies, in hyperbolic (**A** and **C**) and elliptic (**B** and **D**) ranges.

## 5. Analytical expression for the field distribution

From system (S6) we obtain the following relations between coefficients $a_i$ and $b_i$:

$$a_2 = a_1 \frac{(\varepsilon_z q_{ez} - i\varepsilon_2 q)e^{-qk_0 d}}{2\varepsilon_z q_{ez}\cos(q_{ez}k_0 d) + 2\varepsilon_2 q \sin(q_{ez}k_0 d)}$$

$$b_2 = a_1 \frac{(\varepsilon_z q_{ez} + i\varepsilon_2 q)e^{-qk_0 d}}{2\varepsilon_z q_{ez}\cos(q_{ez}k_0 d) + 2\varepsilon_2 q \sin(q_{ez}k_0 d)} \tag{S20}$$

$$b_3 = a_1 \frac{\varepsilon_z q_{ez} e^{-qk_0 d}}{\varepsilon_z q_{ez}\cos(q_{ez}k_0 d) + \varepsilon_2 q \sin(q_{ez}k_0 d)}$$

Therefore, using Eqs. (S5) and (S20) we calculate the field distribution of the plane wave propagating along the biaxial slab. Note, that according to Eq. (S5), the in-plane field distribution can be written as follows

$$\vec{E} = \vec{E}_0(z)e^{i(k_x x + k_y y)} = \vec{E}_0(z)e^{ik_0(q_x' x + q_y' y) - k_0(q_x'' x + q_y'' y)} \tag{S21}$$

In particular, for the direction $\mathbf{r}$, parallel to $\boldsymbol{q}''$, the field distribution reads

$$\vec{E} = \vec{E}_0(z)e^{ik_0(\boldsymbol{q}'\cdot\mathbf{r})-k_0(\boldsymbol{q}''\cdot\mathbf{r})} = \vec{E}_0(z)e^{ik_0q'r\cos\varphi_{\mathrm{v}}-k_0q''r}, \tag{S22}$$

where $\varphi_{\mathrm{v}}$ is the angle between $\mathbf{q}''_{\mathrm{v}}$ and $\mathbf{q}'_{\mathrm{v}}$.

### 6. Conditions for the surface or volume type of the mode

Here we consider the condition that the mode should satisfy to be surface-like. To do this, the imaginary part of $k_z$ inside the slab should be larger than the real part, therefore, the dispersion relation reads:

$$\tanh(-ik_{ez}d) = -\frac{(\varepsilon_1+\varepsilon_2)\varepsilon_z k_{ez}k_{oz}}{\varepsilon_1\varepsilon_3 k_{oz}^2+\varepsilon_z^2 k_{ez}^2} \tag{S23}$$

Neglecting the losses, we notice, that $-ik_{ez}$ is real and positive, consequently, $\tanh(-ik_{ez}d) \in [0, 1]$. As $k_{oz}^2$, $k_{ez}^2$, $k_{ez}k_{oz}$ are negative, and $\varepsilon_1$, $\varepsilon_2$ are positive, to satisfy Eq. (S23) $\varepsilon_z$ should be negative, and the following inequality must be fulfilled:

$$(\varepsilon_1+\varepsilon_2)\varepsilon_z k_{ez}k_{oz} < -(\varepsilon_1\varepsilon_3 k_{oz}^2+\varepsilon_z^2 k_{ez}^2) \tag{S24}$$

On the other hand, $k_{ez} = \sqrt{-\frac{\varepsilon_x}{\varepsilon_z}k_x^2 - \frac{\varepsilon_y}{\varepsilon_z}k_y^2}$. For $k_{ez}$ to be imaginary, one of the in-plane dielectric permittivity tensor components also must be negative. Substituting the expressions for $k_{oz}$ and $k_{ez}$ into Eq. (S24), we obtain:

$$(\varepsilon_1+\varepsilon_2)\sqrt{\varepsilon_x\varepsilon_z\cos^2\alpha+\varepsilon_y\varepsilon_z\sin^2\alpha} < \varepsilon_1\varepsilon_3+\varepsilon_x\varepsilon_z\cos^2\alpha+\varepsilon_y\varepsilon_z\sin^2\alpha \tag{S25}$$

Thus, to observe the propagation of the surface mode along the $j$-direction ($j = x, y$), the following conditions must be fulfilled: 1) $\varepsilon_z < 0$; 2) $\varepsilon_j < 0$; 3) $(\varepsilon_1+\varepsilon_2)\sqrt{\varepsilon_j\varepsilon_z} < \varepsilon_1\varepsilon_3+\varepsilon_j\varepsilon_z$. Usually, all components of the dielectric permittivity of the biaxial slab are high, then, if the first and the second conditions are fulfilled, the third condition almost always will be satisfied.

### 7. Ultrathin-slab limit

Here we simplify Eq. (3) of the main text assuming that $\alpha_j \ll 1$, which is usually satisfied for the thin slabs of van der Waals crystals. Firstly, let us consider the case of high-$k$, under this condition, we assume $q_{1z} \approx q_{2z} \approx iq$, therefore the last term in the first bracket dominates, and Eq. (3) of the main text transforms to the following:

$$\alpha_x q_x^2+\alpha_y q_y^2+q\,\frac{\varepsilon_1+\varepsilon_2}{2i} = \frac{2q_x^2q_y^2(\alpha_x-\alpha_y)^2}{q^3} \tag{S26}$$

As a result, we can neglect the right-hand part of the equation, and the left-hand side immediately brings us to Eq. (S10), defining the dispersion relation of the TM modes under the high-$k$ and ultrathin slab approximations applied simultaneously.

Secondly, In the case of $q$ is close to 1 ($|q-1| \ll 1$), $q_{1z}$ and $q_{2z}$ become small, so the last term in the first bracket dominates, and Eq. (3) transforms into the following:

$$\alpha_x q_y^2 + \alpha_y q_x^2 + \frac{q_x^2+q_y^2}{2}(q_{1z}+q_{2z}) = \frac{2q_x^2 q_y^2(\alpha_x-\alpha_y)^2}{q^2\left(\frac{\varepsilon_1}{q_{1z}}+\frac{\varepsilon_2}{q_{2z}}\right)} \quad \text{(S27)}$$

The right-hand side of this equation is negligible as being the second order on $\alpha_j$ ($j = x, y$), which brings us to the simplified dispersion relation of the TE modes:

$$\alpha_x q_y^2 + \alpha_y q_x^2 + \frac{q_x^2+q_y^2}{2}(q_{1z}+q_{2z}) = 0 \quad \text{(S28)}$$

Finally, the last possible case is $q\sim1$, but not close to 1 ($|q-1|\sim1$), the last terms in both brackets dominate and the simplified Eq. (3) reads:

$$(q_{1z}+q_{2z})\left(\frac{\varepsilon_1}{q_{1z}}+\frac{\varepsilon_2}{q_{2z}}\right) = 0, \quad \text{(S29)}$$

which does not have a solution if both dielectric permittivities of the substrate and superstrate are positive. Thus, for a low 2D conductivity, $\alpha_j \ll 1$, the right-hand side of Eq. (3) remains negligible, guaranteeing the decoupling of the TE and TM modes, which brings us to the system (4).

To analyze the existence of the solution of Eq. (S28) we rewrite it in the following form:

$$k_0 d\varepsilon_x q_y^2 + k_0 d\varepsilon_y q_x^2 = q^2\left(\sqrt{q^2-\varepsilon_1}+\sqrt{q^2-\varepsilon_2}\right) \quad \text{(S30)}$$

The existence of the TE modes is equivalent to $q^2 > 0, \varepsilon_1, \varepsilon_3$, which means that the r.h.s. of Eq. (S30) is positive. Therefore, the l.h.s. of Eq. (S30) is also positive, then, the TE mode in elliptic ranges requires $\varepsilon_x, \varepsilon_y > 0$. In hyperbolic ranges, that is, when the in-plane components of the dielectric permittivity tensor have different signs, the terms in the l.h.s. of Eq. (S30) have different signs. Without loss of generality, we assume that $\varepsilon_x > 0$ and $\varepsilon_y < 0$, then, to satisfy the condition $k_0 d\varepsilon_x q_y^2 + k_0 d\varepsilon_y q_x^2 > 0$, that is, $\varepsilon_x q_y^2 > -\varepsilon_y q_x^2$, the mode should propagate in a sector around the $y$-axis, which means that the electric field should be in a sector around the $x$-axis, corresponding to the positive main component of the dielectric permittivity tensor. Noticeably, as will be shown below, the sector of the propagation of the TM modes is centered along the same axis as the sector of the TE modes. Indeed, the ultrathin-slab limit, the dispersion relation of the TM modes reads:

$$\alpha_x q_x^2 + \alpha_y q_y^2 + \frac{q_x^2+q_y^2}{2}\left(\frac{\varepsilon_1}{q_{1z}}+\frac{\varepsilon_2}{q_{2z}}\right) = 0 \quad \text{(S31)}$$

This equation can be rewritten as follows:

$$k_0 d\left(\varepsilon_x q_x^2 + \varepsilon_y q_y^2\right) = -q(\varepsilon_1+\varepsilon_2). \quad \text{(S32)}$$

To satisfy this equation, the condition $\varepsilon_x q_x^2 + \varepsilon_y q_y^2 < 0$ should be fulfilled, therefore, when $\varepsilon_x > 0$ and $\varepsilon_y < 0$, the TM modes, as well as the TE modes, propagate in a sector centered along the $y$-axis.